# The Medium Redshift Clusters CL0017-20 and CL0500-24

L. Infante[*,1], P. Fouqué[1,2], G. Hertling[1], M.J. Way[1], E. Giraud[**,3], and H. Quintana[***,1]

[1] Grupo de Astrofísica, Facultad de Física, P. Universidad Católica de Chile, Casilla 104, Santiago 22, Chile
[2] Observatoire de Paris-Meudon F-92195 Meudon Principal CEDEX, France
[3] Centre de Physique Théorique - C.N.R.S., Luminy Case 907, F-13288 Marseille Cedex 9, France
Internet: linfante@astro.puc.cl

**Abstract.** We present magnitudes and redshifts of galaxies in the cluster of galaxies CL0017-20 and in the double cluster CL0500-24. Luminosity functions and velocity distributions have been derived. CL0017-20 possesses a virial mass of $2.9 \times 10^{14}$ M$_\odot$ and a $V$ band luminosity of $2.3 \pm 0.5 \times 10^{12}$ L$_\odot$ ($M/L_V = 127$), much of it in a very compact core of size 113 kpc. Our new velocities and photometry of CL0500-24 confirm its double nature. Our group algorithm clearly detects two subconcentrations with masses 1.9 and $2.2 \times 10^{14}$ M$_\odot$. Its total $V$ band luminosity amounts to $2.7 \pm 0.8 \times 10^{12}$ L$_\odot$ ($M/L_V$ of 157). Since the crossing and collapse times in both clusters are smaller than the age of the Universe, they appear bound and virialized. It is surprising that although $M/L$ is small in these clusters, giant arcs have been detected in both.

## 1. Introduction

The largest collapsed mass aggregates known in the Universe, clusters of galaxies, provide both a direct estimate of the amplitude of the spectrum of fluctuations in the Universe and an estimate of $\Omega$. Current estimates of the mass-to-light ratios in medium z clusters, $\sim 70 - 150$ M$_\odot$/L$_\odot$ (Mellier et al. 1988, Infante et al. 1992 (but see §5), Sharples et al. 1985), are well below what is required to close the Universe. X-ray luminosities and motion of galaxies have been used to study the gravitational field around cluster cores and, thus, estimate their masses. An independent way to obtain the total mass in clusters is now available using gravitational lensing (Wu and Hammer, 1993). In fact, gravitational arcs have become an additional test to probe not only the existence and amount of dark matter in clusters but also how this matter is distributed. (For a review see Tyson 1992.). CL0017-20 and CL0500-24 are two medium z rich clusters. Although both clusters have gravitational arcs they show a very distinct dynamical structure.

CL0017-20 (Fig. ??), a rich cluster at a redshift of 0.272, has turn out to be quite an interesting case. It was first discovered on deep CFHT prime focus plates by Infante et al. (1986). It is located near the South Galactic Pole and contains a giant galaxy at its centre, probably a cD galaxy. This galaxy is surrounded by several smaller galaxies in a disk like configuration all embedded in what seems to be a common, extremely compact, halo of diameter $\sim 113$ kpc ($q_0 = 0.5, H_0 = 60$ km s$^{-1}$ Mpc$^{-1}$, adopted cosmology hereafter) and total absolute magnitude in V of $-24.1$. Luppino et al. (1991) discuss the photometry of a similar cluster, CL1358+6245 at $z = 0.32$.

The brightest members of this cluster are clearly very red (i.e. $B - V = 0.9$, at $z = 0$) as would be expected for a cluster dominated by E/S0 galaxies. However, a significant blue population of galaxies (with $(B - V) < 0.7$, which

*Send offprint requests to*: L. Infante
[*] Visiting astronomer, Las Campanas Observatory of the Carnegie Institution of Washington and European Southern Observatory, La Silla
[**] Visiting astronomer, European Southern Observatory, La Silla
[***] Visiting astronomer, Las Campanas Observatory of the Carnegie Institution of Washington

would correspond to later than Sab spirals) is also found, consistent with the findings of Butcher and Oemler (1984) of a higher fraction of blue to red galaxies in medium redshift clusters as compared to low z clusters.

Infante, Giraud and Triay (1991) report an arc-like feature on deep images of CL0017-20 acquired with EMMI on the NTT at La Silla on a non-photometric night. Although during the NTT observations the seeing was 0.9", poor for NTT standards, the arc-like feature is quite conspicuous in all the frames. The arc is significantly bluer than the red cluster galaxies.

The galaxy cluster CL0500-24 (Fig. ??) (CL0500-24 in Giraud 1988) is a highly contrasted rich cluster at a redshift of 0.322. It presents a giant arc, rather straight (Giraud 1988, Wambsgauss et al. 1989), and a Butcher-Oemler effect (Giraud 1990). Spectra of the arc have been obtained confirming the gravitational lens interpretation of an object at $z \approx 0.724$ (Giraud 1993). Models for the arc require a multistructured lensing cluster. In this paper we obtain velocities for a few additional cluster members, confirming the double nature of the cluster.

We also present multicolour photometry and spectroscopic data of CL0017-20 and CL0500-24. In §2 observations and data reduction are presented. The results are shown in §3. Luminosity functions and mass-to-light ratios are discussed in §4 and §5, respectively. Conclusions are presented in §6.

## 2. Observations and Data Reduction

### 2.1. Photometry

We have carried out CCD observations of CL0017-20 using a variety of telescopes and filter combinations. Since the discovery of the cluster we have acquired CCD frames with the 40" telescope at Las Campanas (L.C.O.) in April 1987, with the C.T.I.O. 4m telescope in 1989, with the 2.2m telescope at La Silla Observatory (E.S.O.) in 1991 and with the 40" telescope at L.C.O. in 1991. In the present work we report results from our best frames obtained in photometric nights.

The 1987 L.C.O. observations were carried out with a RCA back-illuminated thinned $317 \times 512$ pixel CCD chip on the 40" telescope, f/7 Cassegrain focus. The scale on the image plane is 0.87 arcsec/pixel with a total angular extension of $4.5 \times 7.3$ arcmin$^2$. Three frames in Gunn $r$ and two frames in Gunn $g$ filters were exposed for 1200 seconds each. Dark, bias and flat frames were also acquired throughout the night. All frames were dark and bias subtracted, trimmed and flat-fielded using standard techniques. We then applied a median filter to the three $r$ frames and averaged the two $g$ frames. Residual cosmic rays were then removed using a special purpose delta function analyzer algorithm. The seeing during the observations was typically between 1 and 1.5 arcsec FWHM.

Photometric standards were observed regularly each night. These included standards from Thuan and Gunn (1976), Kent (1985) and galaxies observed in the field by Infante et al. (1986). The photographic $J$ and $F$ in Infante et al. were tranformed to $g$ and $r$ by combining Kent (1985) and Infante et al. relations,

$$g = J - 0.19 - 0.40\,(J - r_{ph}) \tag{1}$$
$$g - r = -0.53 + 0.63\,(J - r_{ph}) \tag{2}$$

where $J$ and $r_{ph}$ are photographic magnitudes. 9 galaxies, marked in capital letters in Fig. ??, were used to determine the instrumental to standard transformation coefficients. The adopted transformation is as follows:

$$g_i = g - 29.41_{(\pm 0.042)} - 0.30_{(\pm 0.049)}(g - r) \tag{3}$$
$$r_i = r - 29.45_{(\pm 0.104)} + 0.16_{(\pm 0.121)}(g - r) \tag{4}$$

where $g_i$ and $r_i$ are instrumental magnitudes, $g$, $r$, and $(g - r)$ are standard magnitudes and colours. The magnitude errors range from $\pm 0.15$ mag at $g = 23$ to $\leq \pm 0.07$ mag at $g = 18$, and from $\pm 0.15$ mag at $r = 22$ to $\leq \pm 0.07$ mag at $r = 17$.

The Gunn $i$ band magnitudes have been obtained in two steps. We first used a 40" L.C.O. Tek $2048 \times 2048$ CCD $i$ ($20' \times 20'$ field) image, observed during a good photometric night in 1991 and calibrated using Landolt $I$ and $V$ standards, to obtain photometry of 21 objects (stars and galaxies) in the CL0017-20 field. We then used those 21 objects to calibrate a much deeper $I$ (Kron-Cousins) image obtained with the 4m telescope at C.T.I.O., TI $393 \times 398$ (kindly made available by C.J. Pritchet). Since $I$ Landolt is shifted only by a constant ($\sim 0.7$) with respect to Gunn $i$, and not by a colour term, we transformed our $I$ Landolt magnitudes to Gunn $i$.

On February, 1993, $V$ and $I$ bands images of CL0500-24 were obtained with the Tek $2048 \times 2048$ CCD mounted at the Cassegrain focus of the 40" L.C.O. telescope. Three 300 sec frames were acquired for each bandpass. A median of the frames, processed using the standard procedure of bias subtraction and twilight flat-fielding, was taken. Landolt (1992) standards were used to calibrate the photometry. The adopted extinction and transformation coefficients were as follows:

$$V = V_i - 21.20 - 0.17\,X + 0.042\,(V - I) \tag{5}$$
$$I = I_i - 20.76 - 0.07\,X + 0.026\,(V - I) \tag{6}$$

where $V_i$ and $I_i$ are instrumental magnitudes and $X$ is the air mass.

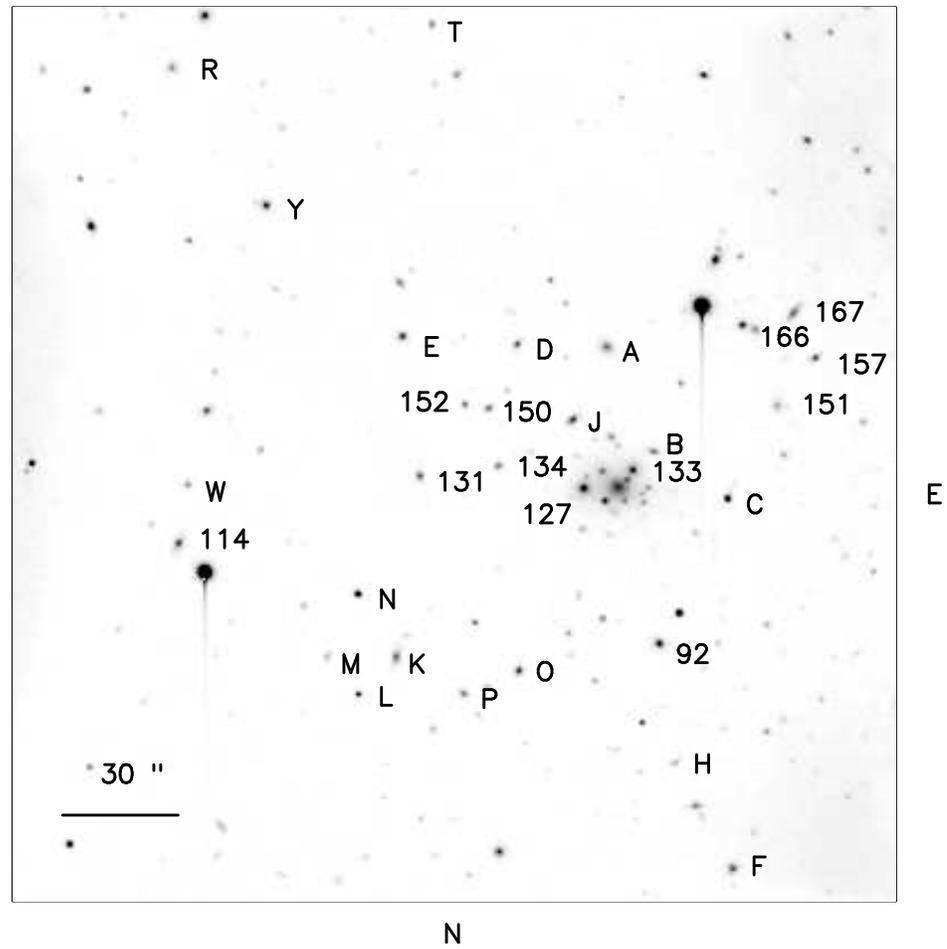

**Fig. 1.** Central region of CL0017-20. Galaxies marked in capital letters were used as standards. This is a grey scale plot of a C.T.I.O. 4m I CCD image.

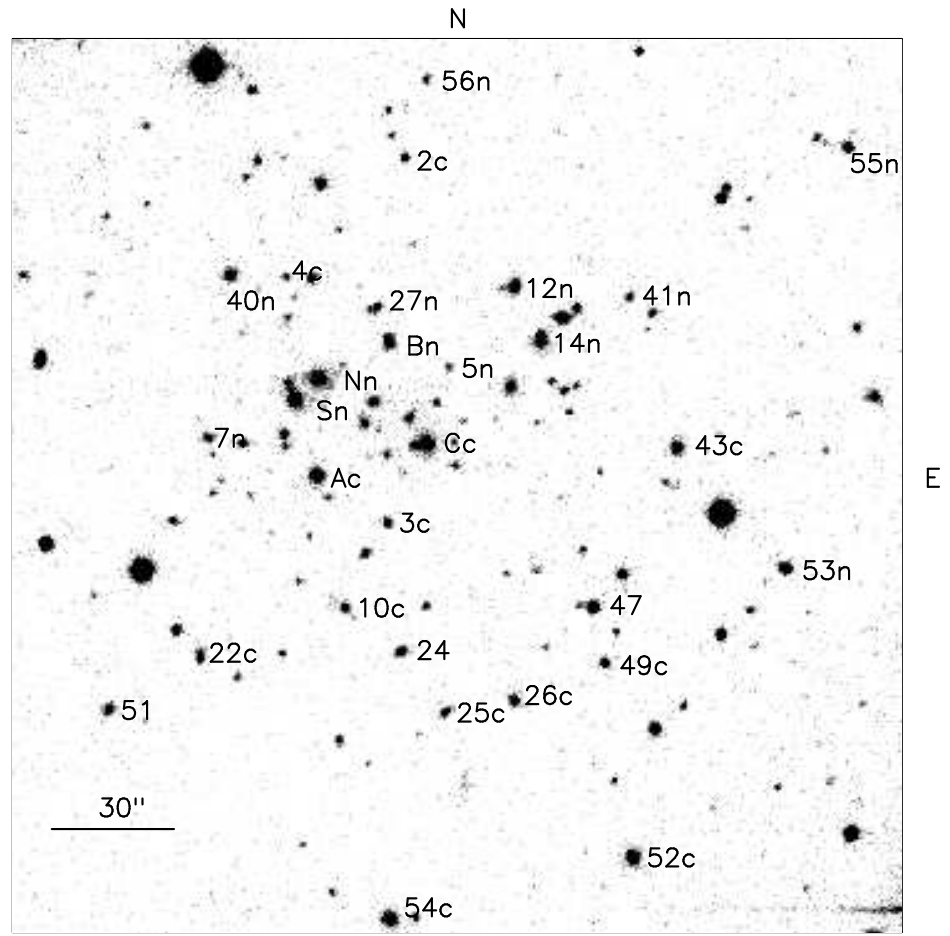

**Fig. 2.** Central region in CL0500-24. Galaxies marked with a **n** belong to the northern concentration (centre galaxy **N**) and galaxies marked with a **c** belong to the southern concentration (centre galaxy **C**). This is a grey scale plot of a L.C.O. 40" V image.

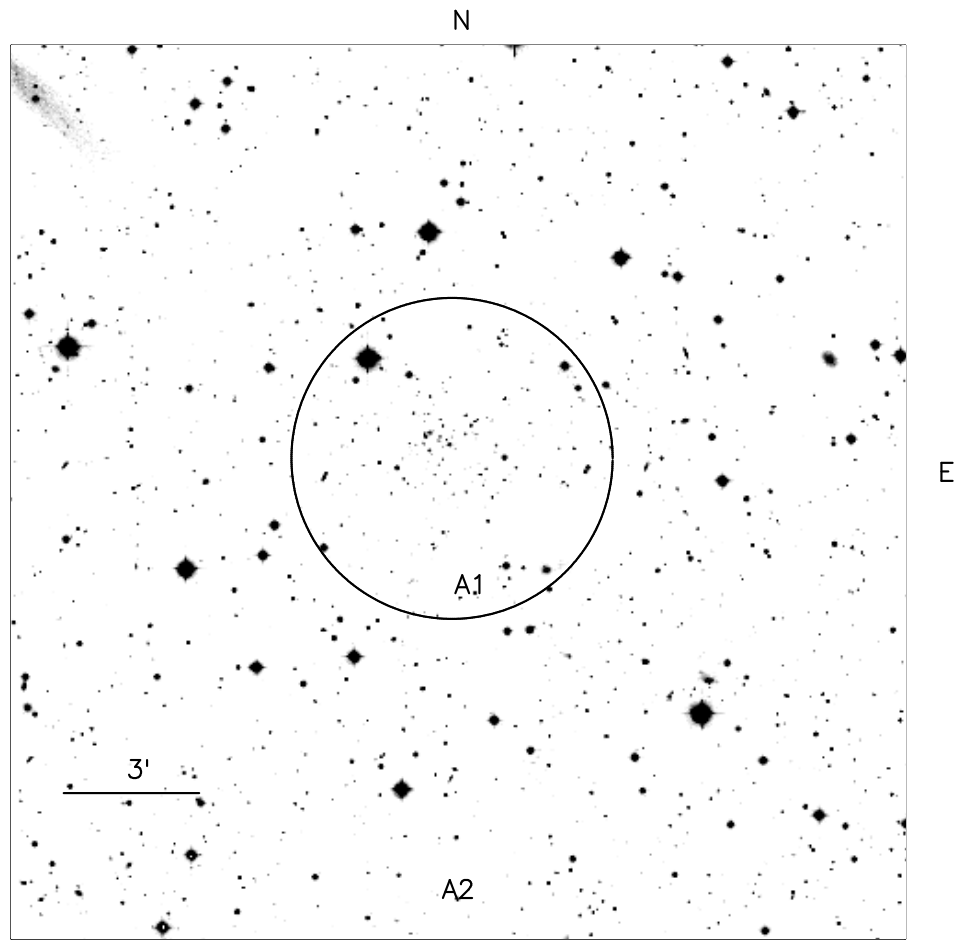

**Fig. 3.** CL0500-24. The central area **A1** contains the cluster core. **A2** is the area from which the background contribution is calculated.

Image detection, photometry and classification were performed using the methods described in Kron (1980), Infante (1987) and Infante and Pritchet (1992). "Total" magnitudes were calculated by integrating the intensity within elliptical apertures. Care was taken to choose apertures for each object which correspond in both passbands to the same physical dimensions.

### 2.2. Spectroscopy

Multislit observations of 5 fields of galaxies were carried out on three non-photometric nights in December 1990, using EFOSC 1 at the Cassegrain focus of the ESO 3.6 meter telescope. The grism used gave a dispersion of 230Å mm$^{-1}$. We used CCD No. 11 which has $1024 \times 640$ pixels, and a pixel size of 15 microns. The CCD was binned 2-by-2 giving a projected pixel size of 0.68".

Aperture masks were made for each field consisting of 9 to 14 slits with lengths of $\sim 10$" and widths of $\sim 2$". The spectra were reduced twice to the point of assigning wavelength. An initial reduction was performed using IHAP software at La Silla. A more methodical reduction was later performed with IRAF in Santiago, Chile. Both methods were in agreement and we found no systematic trends between the two reductions. The data were bias subtracted and flat-fielded. Sky removal was done by subtraction of sections adjacent to the spectrum in each slit. Wavelength solutions were calculated for Helium-Argon arc lamp exposures. The rms fit of the arc lines in IRAF was 0.5Å pixel$^{-1}$ while using a $4^{th}$ order polynomial to fit approximately 25 lines. The wavelength range span from approximately 3000Å to 7000Å and we obtained 6.9Å pixel$^{-1}$. The resolution was 13.5Å.

Cosmic rays were removed via the COSMICRAY routine in IRAF. The reduction was actually done twice within IRAF, both with and without the removal of cosmic rays. The COSMICRAY algorithm in IRAF can remove both cosmic rays, real emission lines, and can distort deep adjacent absorption features such as the Ca II H and K lines so one has to adjust the task parameters such that only cosmic rays are removed. The program uses the ratio of the flux of the cosmic ray candidate to neighboring pixels to determine its likelihood.

The determination of redshift required two methods. One was a line by line absorption and emission identification and the other via RV, the Fourier cross-correlation routine in IRAF. For the template spectrum we used a stellar spectrum that appeared in one of the field slits. This spectrum was first wavelength calibrated and then used in the RV task. We also used another galaxy template from another observing run for a check on systematics and none were found. Prior to cross-correlation, all real emission and residual night sky emission line features were removed through interpolation and a $10^{th}$ order Legendre polynomial was fit to the spectra for continuum removal. A simple parabola was fit to the correlation peak. For spectra whose Tonry & Davis $R$ value was below 3.0, a comparison was made with absorption line identifications. If no absorption lines could be identified in the spectra to confirm the RV result, the velocity was discarded. The Tonry & Davis $R$ value is a measure of the quality of the fitted correlation peak which is defined as the ratio of the true peak height to the average height of the peaks in the antisymmetric component of the cross-correlation function, see Tonry & Davis (1979).

The velocity of the central core of CL0017-20 was measured from a spectrum obtained in 1987 with the 2.5m L.C.O. Observatory 2D-Frutti spectrograph. The measured velocity is $81435 \pm 68$ km s$^{-1}$ (z=0.272) (Quintana & Ramirez 1993).

## 3. Results

### 3.1. Magnitudes and colours

#### 3.1.1. CL0017-20

We have obtained $g$ and $r$ magnitudes for 213 galaxies, $i$ for 80 galaxies and velocities for 26 galaxies. Table ?? shows $g$, $(g$-$r)$, $(r$-$i)$, velocity and velocity uncertainty for galaxies identified in Fig. ??. We have chosen not to include the Table containing all the magnitudes because of its length. Instead, we present a colour-magnitude diagram and a colour-colour diagram of galaxies in the field of CL0017-20. (See Fig. ?? and ??.)

Due to the central crowding our analysis has been performed using galaxies outside the central 26.1 arcsec circle. Plots also show the data for galaxies external to the central circle. In order to estimate the magnitude and colour of the central core, which probably dominates the potential well of the cluster, we have performed concentric aperture photometry centered on the cluster luminosity-weighted centre. The results are shown in Table ??. We adopt $g = 17.54$ and $(g-r) = 1.25$ as the central (diameter = 26.1 arcsec = 112.7 kpc) magnitude and colour respectively. This nuclear light dominates the light of the cluster. It is interesting to note that there is no colour gradient in the core. Luppino et al. 1991 find a very small gradient in $(V-R)$ index in CL1358+6245. However, they conclude that the effect is not significant enough.

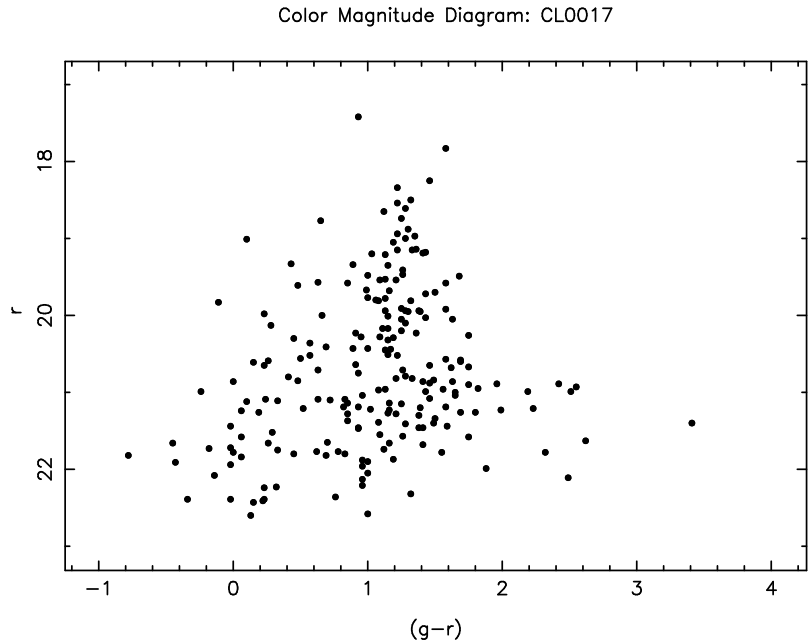

**Fig. 4.** Colour-Magnitude diagram of galaxies in the field of CL0017-20. The photometric uncertainties run from ±0.15 mag at $g = 23$ to $\leq \pm 0.07$ mag at $g = 18$, and from ±0.15 mag at $r = 22$ to $\leq \pm 0.07$ mag at $r = 17$.

3.1.2. CL0500-24

$V$ and $I$ magnitudes for galaxies in areas **A1** and **A2** as shown in Fig. ?? are presented. The central area **A1** contains the cluster core. Its angular size, $\theta_{A1} = 7.1$ arcmin, corresponds to a linear size $D_{A1} = 2.04$ Mpc ($q_o = 0.5, H_o = 60$ km s$^{-1}$ Mpc$^{-1}$). The total useful solid angle of **A1** is 0.01017 deg$^2$. **A2** is the area from which the background contribution is calculated. The total useful solid angle of **A2** is 0.09679 deg$^2$. Galaxies have been counted as a function of magnitude in **A2**. If a slope of 0.4 is assumed (a formal least-squares fit to the counts gives a slope of 0.39 with an rms of 0.04) a fit to the number counts gives $\log N = 0.4V - 5.37$ [N/deg$^2$/mag], which, after the appropriate passband transformation $J = V + 1.034\ (B - V)$, is consistent with the South Galactic Pole **field** counts in Infante et al. 1986, $\log N = 0.4V - 4.99$ [N/deg$^2$/mag]. Thus, the cluster contribution in **A2** is negligible.

Table ?? shows $V$, $(V-I)$, velocity and velocity uncertainty for galaxies identified in Fig. ??. Figure ?? is a colour-magnitude diagram for CL0500-24. Filled circles represent galaxies in **A1** and open circles correspond to galaxies in **A2**. The cluster mean colour is $< V - I > \approx 1.8$ while the mean colour for the field galaxies is $< V - I > \approx 1.4$. It is interesting to note that the filled circles, which would correspond to the blue cluster population i.e. Butcher-Oemler objects, are also at a mean colour of 1.4. Clearly, most of those blue galaxies are field galaxies and not B-O galaxies.

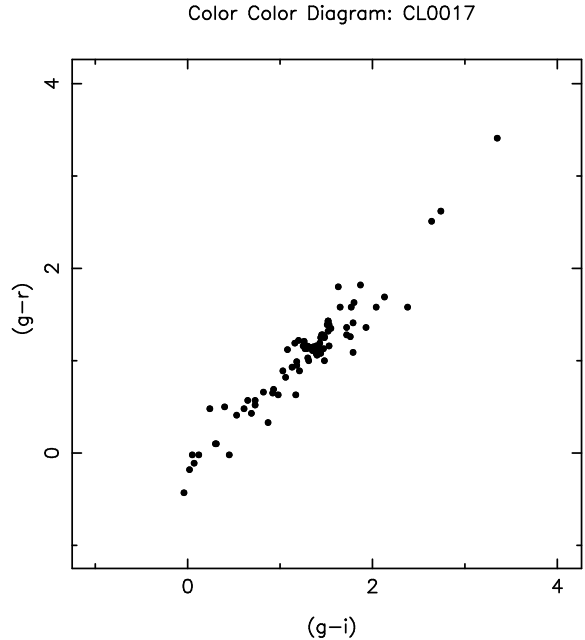

**Fig. 5.** Colour-Colour diagram of galaxies in the field of CL0017-20. The photometric uncertainties run from $\pm 0.15$ mag at $g = 23$ to $\leq \pm 0.07$ mag at $g = 18$, and from $\pm 0.15$ mag at $r = 22$ to $\leq \pm 0.07$ mag at $r = 17$.

*3.2. Velocities*

In Fig. ?? we show the distribution of velocities in CL0017-20. After clipping out 3 $\sigma$ outliers 22 out of 26 galaxies remained. A Kolmogorov-Smirnov test which compares our distribution of velocities to a gaussian distribution returned a significance level larger than 95%. The observed line-of-sight velocity dispersion and mean recession velocity are $\sigma = 1510 \pm 195$ km s$^{-1}$ and $v = 81579$ km s$^{-1}$ respectively. Uncertainties of velocity dispersions are calculated using the Jackknife method (Beers et al. 1990).

Table ?? contains 29 velocities; 25 of them have been taken from Giraud 1990. We have added 4 more velocities. These are for objects 2c, 4c, 5n and 7n identified in Fig. ??. Objects in Table ?? and in Fig. ?? are labeled according to the subconcentration they belong to, i.e. $n$ or $c$. There are 13 objects in each subconcentration. Two of the remaining objects are foreground and one is background. In Fig. ?? we show the distribution of velocities in CL0500-24. The mean velocity and observed one dimensional velocity dispersion of the northern concentration are 97987 km s$^{-1}$ and $1152 \pm 214$ km s$^{-1}$ respectively, while the southern concentration, centered on galaxy C, has a mean velocity of 94817 km s$^{-1}$ and a one dimensional velocity dispersion of $917 \pm 208$ km s$^{-1}$. The mean velocity of the 26 cluster members is 96402 km s$^{-1}$ and the observed one dimensional velocity dispersion is $1911 \pm 216$ km s$^{-1}$. These two values compare very well with Giraud 1990 values. The effect of the 4 additional velocities is to enhance both peaks in Fig. ??.

## 4. Luminosity function and richness

*4.1. Background correction*

Before fitting a luminosity function to the counts of galaxies in the cluster area, we have to remove the background contamination. For this purpose, we start with the field galaxy counts in two colours, published by Infante et al. (1986). From photographic plates in $J$ and $F$, Infante determined the following differential galaxy counts per square

**Table 1.** CL0017-20: Magnitudes and Velocities

| Id. | $g$ | $(g\text{-}r)$ | $(r\text{-}i)$ | $Vel$ km s$^{-1}$ | $\Delta Vel$ km s$^{-1}$ |
|---|---|---|---|---|---|
| Core | 17.54 | 1.25 | 0.47 | 81440 | 68 |
| L | 22.27 | 1.69 | 0.44 | 91690 | 46 |
| P | 21.29 | 1.11 | 0.23 | 81900 | 113 |
| O | 20.77 | 1.00 | 0.31 | 83900 | 204 |
| M | 22.07 | 1.28 | 0.17 | 79900 | 149 |
| 92 | 20.28 | 1.28 | 0.18 | 79180 | 116 |
| N[1] | 21.72 | 0.63 | 1.17 | 82100 | 113 |
| 114 | 20.23 | 1.03 | 0.27 | 83150 | 179 |
| C | 21.16 | 1.58 | 0.46 | 91470 | 162 |
| 127 | 20.18 | 1.30 |  | 83670 | 87 |
| W | 21.66 | 1.15 | 0.16 | 80210 | 115 |
| 131 | 20.89 | 1.08 | 0.31 | 81670 | 135 |
| 133 | 20.37 | 1.22 |  | 82790 | 156 |
| 134 | 21.48 | 1.19 | 0.34 | 80220 | 101 |
| B[2] | 21.46 | 1.43 | 0.06 | 79070 | 67 |
| J | 20.84 | 1.16 | 0.37 | 80750 | 85 |
| 150 | 21.32 | 1.15 | 0.21 | 81160 | 143 |
| 151 | 21.34 | 1.39 | 0.14 | 154410 | 62 |
| 152 | 21.47 | 1.15 | 0.16 | 81620 | 150 |
| 157 | 21.30 | 1.25 | 0.23 | 81580 | 57 |
| D | 20.66 | 0.66 | 0.16 | 84490 | 128 |
| A | 19.76 | 0.43 | 0.26 | 65740 | 124 |
| E | 20.73 | 1.26 | 0.22 | 83050 | 105 |
| 166[2] | 20.34 | 1.13 | 0.16 | 82230 | 75 |
| 167 | 20.20 | 0.63 | 0.63 | 80040 | 269 |
| Y | 20.23 | 0.89 | 0.32 | 80650 | 101 |

[1] possible cosmic ray in the $i$ image
[2] L.C.O. 1991 photometry

**Table 2.** CL0017-20: Aperture Photometry of the Core

| $r_i$ kpc | $r_o$ kpc | $g$ | $(g\text{-}r)$ | $(r\text{-}i)$ | $\Sigma_g$ mag/arcsec$^2$ | $\Sigma_r$ mag/arcsec$^2$ |
|---|---|---|---|---|---|---|
| 0 | 18.8 | 18.82 | 1.25 | 0.39 | 23.25 | 21.99 |
| 18.8 | 37.6 | 17.84 | 1.22 | 0.44 | 23.79 | 22.57 |
| 37.6 | 56.4 | 17.54 | 1.25 | 0.47 | 24.36 | 23.11 |

\* $r_i$ and $r_o$ are the inner and outer ring boundaries.
\* Adopted cosmology: $q_\circ = 0.5$, $H_\circ = 60$ km s$^{-1}$ Mpc$^{-1}$.

degree per magnitude:

$$\log N = 0.4\,J - 5.40 \quad (7)$$

$$\log N = 0.35\,r_{ph} - 3.72 \quad (8)$$

The second equation was used directly to correct $r$ counts in CL0017-20, assuming that photographic and CCD Gunn $r$ magnitudes are on the same scale. As the CCD cluster area is 0.00725 square degree, the relevant equation is $\log N = 0.35\,r - 5.86$.

A slightly more precise method was used to correct $g$ counts in CL0017-20, as the $J$ counts were available. Since the mean colour of the galaxies in the magnitude range 18 to 22 is approximately constant ($<J-r>=1.5$), a simple colour conversion was done between $J$ and $g$, using relevant equations from Kent (1985) and Infante et al. (1986),

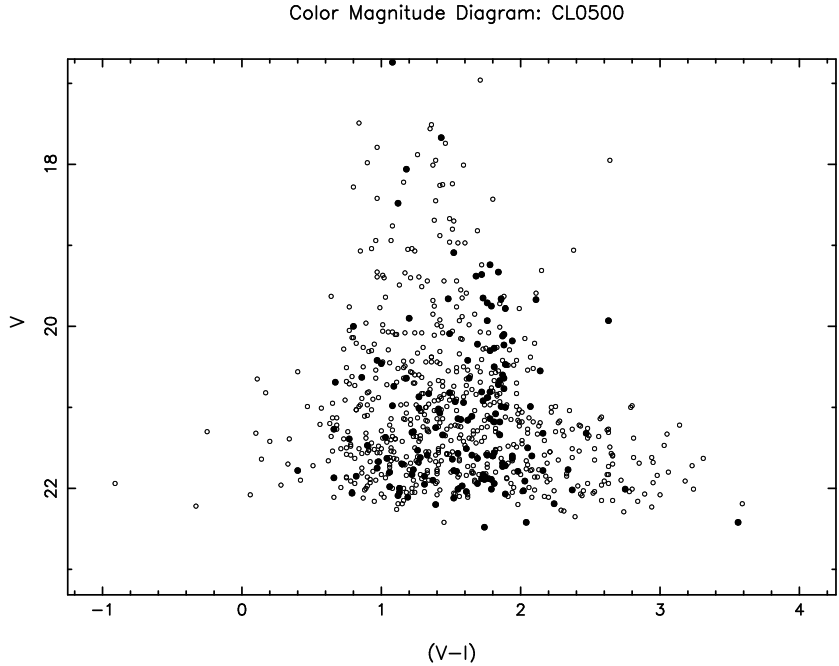

**Fig. 6.** Colour-Magnitude diagram of galaxies in the field of CL0500-24. Open circles are objects in Area **A2**. Filled circles are objects in the central region **A1** in Fig. **??**. The photometric uncertainties run from ±0.1 mag at $V = 21$ to $\leq \pm 0.05$ mag at $V = 18$, and from ±0.15 mag at $i = 22$ to $\leq \pm 0.05$ mag at $i = 17$.

namely:

$$g = V - 0.19 + 0.41\,(B - V) \tag{9}$$
$$J - r = 1.56\,(B - V) \tag{10}$$
$$J = B + 0.03\,(B - V), \tag{11}$$

which gives $g = J - 0.19 - 0.40\,(J - r)$, or $g = J - 0.79$ for the mean galaxy colour. Since the $g$ and $J$ bands are close, we assume the same slope (0.4) for both. Applying the conversion equation and counting the number of galaxies in each bin, we get the following result for the CCD cluster area: $\log N = 0.4\,g - 7.11$.

An even more precise method was used to correct $V$ counts in CL0500-24. The CCD area was large compared to the cluster area, thus allowing to count field galaxies in the external area **A2** of the CCD image. We adopt again the slope 0.4 for the $V$ counts. The mean zero-point in the external area **A2** is then $-6.38 \pm 0.02$. Correcting for the ratio between areas **A2** and **A1** (0.09679/0.01017), the adopted equation reads: $\log N = 0.4\,V - 7.36$.

### 4.2. Fit of the Schechter luminosity function

Two algorithms have been used to fit the Schechter luminosity function (hereafter SLF) to the differential galaxy counts in these clusters:

- The first one (hereafter, Alg 1) makes a non-linear least-squares fit to binned data, using the Levenberg-Marquardt method (Numerical Recipes, Press et al. 1988). It allows to derive all three parameters of the SLF ($\alpha$, $m^\star$, $n^\star$), or to fix some of them and derive the others. It returns a $\chi^2$ value to judge the goodness of the fit.

- The second one (hereafter, Alg 2) makes a maximum likelihood fit to unbinned data, according to Schechter & Press (1976). It needs an a priori knowledge of $\alpha$.

**Table 3.** CL0500-24: Magnitudes and Velocities

| Id. | V | (V-I) | Vel. km s$^{-1}$ | $\Delta Vel$ km s$^{-1}$ |
|---|---|---|---|---|
| Ac | 19.78 | 1.89 | 95630 | |
| Bn | 20.10 | 1.88 | 97130 | |
| Cc | 19.33 | 1.85 | 94730 | |
| Nn | 19.24 | 1.79 | 98330 | |
| Sn | 19.09 | 1.52 | 98330 | |
| 2c | 21.11 | 1.65 | 92960 | 120 |
| 3c | 20.99 | 1.86 | 94440 | |
| 4c | 21.72 | 1.89 | 95790 | 92 |
| 5n | 22.00 | 1.13 | 98810 | 137 |
| 7n | 20.82 | 1.49 | 97570 | 116 |
| 10c | 20.81 | 1.78 | 95330 | |
| 12n | 19.93 | 1.76 | 96230 | |
| 14n | 19.71 | 1.76 | 96230 | |
| 22c | 20.64 | 1.18 | 95030 | |
| 24 | 20.46 | 1.00 | 57860 | |
| 25c | 20.83 | 1.34 | 93240 | |
| 26c | 20.50 | 1.81 | 95330 | |
| 27n | 20.87 | 1.27 | 98630 | |
| 40n | 20.18 | 1.94 | 99830 | |
| 41n | 21.25 | 1.40 | 99830 | |
| 43c | 20.08 | 1.48 | 94440 | |
| 47 | 19.90 | 1.20 | 42870 | |
| 49c | 20.93 | 1.53 | 95930 | |
| 51 | 20.55 | 2.14 | 113620 | |
| 52c | 19.66 | 1.87 | 94440 | |
| 53n | 20.12 | 1.87 | 97130 | |
| 54c | 19.65 | 1.73 | 95330 | |
| 55n | 20.42 | 0.97 | 97730 | |
| 56n | 21.37 | 1.03 | 98030 | |

\* The numbers in column 1 (Id.) correspond to Giraud 1990 identification.
\* Velocities were calculated from redshifts given by Giraud 1990 except for galaxies 2c,4c,5n, and 7n. Only those four velocities have errors quoted in the Table. Identical values are due to low precision in the published redshift.
\* n and c in columm Id (1) correspond to the N and C concentrations respectively. Numbers in the Id columm not having n or c are galaxies which do not belong to any concentration.

The correction for background contamination in Alg 1 is straightforward, as data are binned. Equations from §4.1 are used. Alg 2 needs the total number of background galaxies ($n_{\rm bg}$) whose magnitudes fall between the brightest cluster galaxy used for the fit ($m_1$) and the magnitude limit ($m_{\rm lim}$), and the equivalent total magnitude ($m_{\rm bg}$) of these galaxies. Both $m_1$ and $m_{\rm lim}$ have to be determined. $m_1$ is the magnitude of the brightest galaxy after removal of foreground galaxies and of the brightest cluster galaxy if it overhelms the magnitudes of the following ones. This is the case if the central galaxy is a cD, or if various galaxies are clustered at the centre, and the corresponding magnitude encompasses all of them. $m_{\rm lim}$ is first evaluated from a histogram of magnitudes. This is a sensible parameter, because a too deep limit gives a wrong value of $\alpha$, and thus of $m^\star$. Values of $\alpha$ too different from the canonical value $-1.25$ have been discarded, as our counts are not deep enough to determine $\alpha$ confidently.

After several trials, our approach has been the following one:
- the appropriate binning to be used in Alg 1 has been determined by minimizing $\chi^2/\nu$.
- the magnitude limit has been determined using Alg 1, until reaching plausible values of $\alpha$ ($\sim -1.25$).
- **then we have used Alg 2 with $\alpha = -1.25$.**
- we have verified that the resulting values of $m^\star$ and $n^\star$ from both algorithms were compatible.

The adopted values of $m_1$, $m_{\rm lim}$, $n_{\rm bg}$, and $m_{\rm bg}$ for $g$ and $r$ counts in CL0017-20 and $V$ counts in CL0500-24 are given in Table **??**. This Table also gives the a priori values of $\alpha$ determined by Alg 1 (adopting a binning of 0.25 mag for CL0017-20 and 0.3 for CL0500-24), which are all compatible with the adopted value $-1.25$. Then the apparent magnitude of the SLF knee, and the normalization factor, both with their uncertainties, follow.

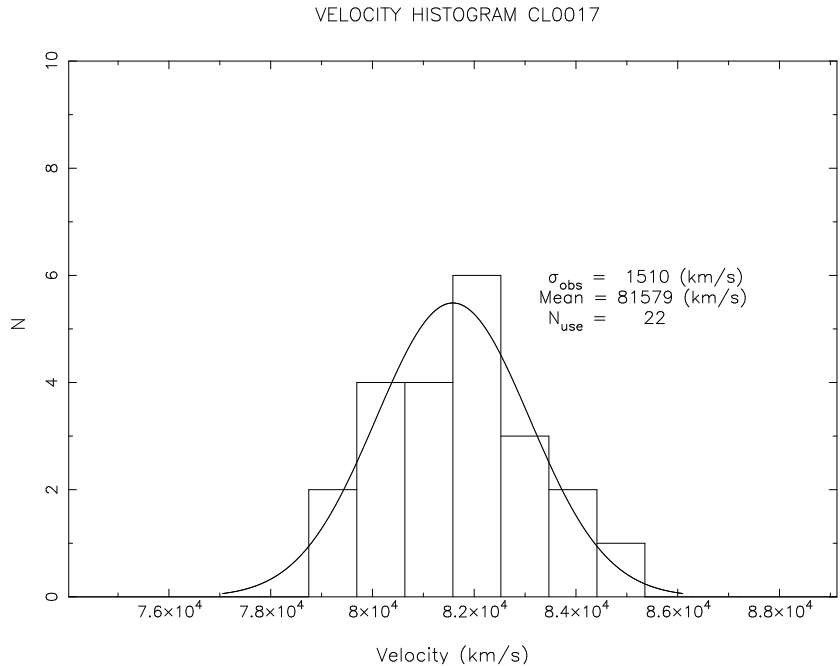

**Fig. 7.** Histogram of velocities in CL0017-20. Galaxies with velocities larger than $3\sigma$ have been removed. 22 out of 26 galaxies remained. The outliers are not in the histogram. We also plot the best gaussian distribution to the points.

**Table 4.** Adopted and resulting parameters of the SLF fit

| Cluster and band | $m_1$ | $m_{\lim}$ | $n_{\rm bg}$ | $m_{\rm bg}$ | $\alpha$ | $m^\star$ | $n^\star$ | $M^\star$ |
|---|---|---|---|---|---|---|---|---|
| CL 0017 $g$ band | 19.11 | 21.5 | 30.9 | 16.83 | $-1.63 \pm 1.07$ | $20.39 \pm 0.24$ | $74 \pm 20$ | $-20.99$ |
| CL 0017 $r$ band | 17.83 | 21 | 38.3 | 15.82 | $-1.29 \pm 0.54$ | $18.99 \pm 0.25$ | $45 \pm 9$ | $-21.91$ |
| CL 0500 $V$ band | 19.09 | 22 | 29.9 | 17.24 | $-1.51 \pm 0.54$ | $19.97 \pm 0.23$ | $52 \pm 10$ | $-21.71$ |

The corresponding values of $m^\star$ given by Alg 1 only differ by 0.01 to 0.02 mag. The fits of the cumulative luminosity function, after correction for background contamination, and forcing $\alpha = -1.25$, are displayed in Fig. ?? for CL0017-20 in $g$ and Fig. ?? for CL0500-24 in $V$. The fit in Fig. ?? does not seem entirely satisfactory at faint levels. As explained above, it is difficult to determine the completeness limit, especially when the true value of $\alpha$ differs from the canonical value of $-1.25$.

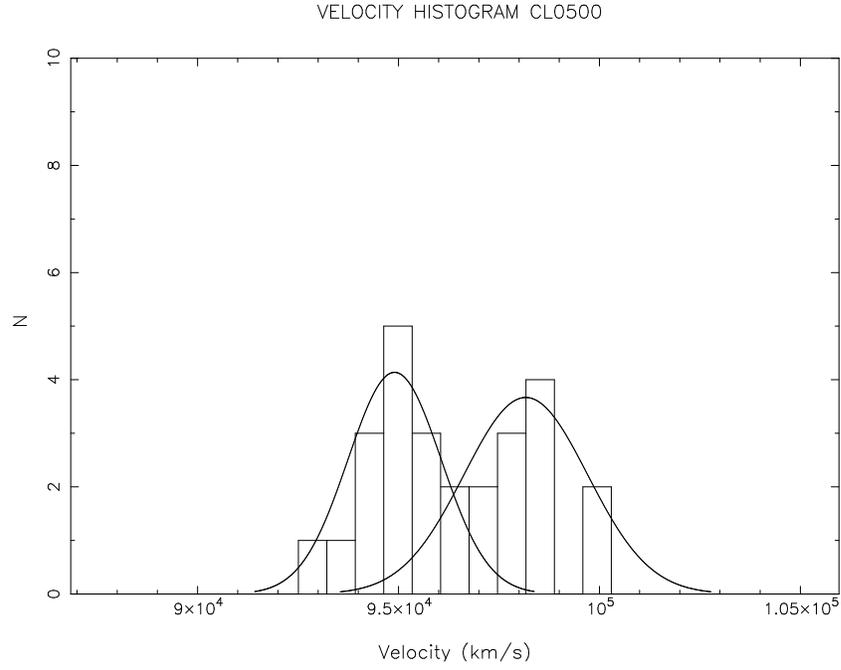

**Fig. 8.** Histogram of velocities in CL0500-24. Galaxies with velocities larger than $3\sigma$ have been removed. 26 out of 29 galaxies remain. We also plot the best gaussian distribution to the points.

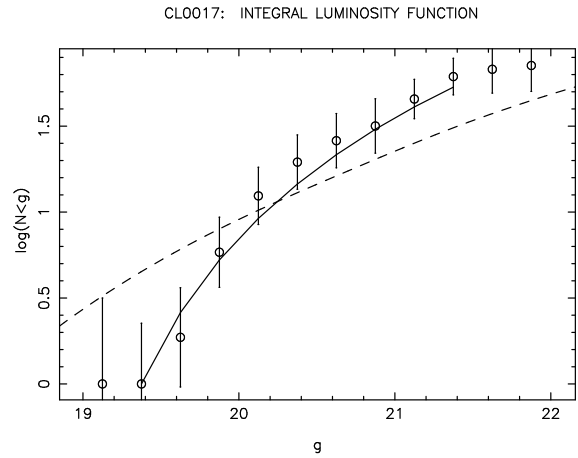

**Fig. 9.** CL0017-20 $g$ Luminosity Function. The dashed line represents the background counts and the solid line is a fit to the Schechter Function. We have adopted 21.5 as the completeness limiting magnitude. The error bars represent $\sqrt{N}$ noise

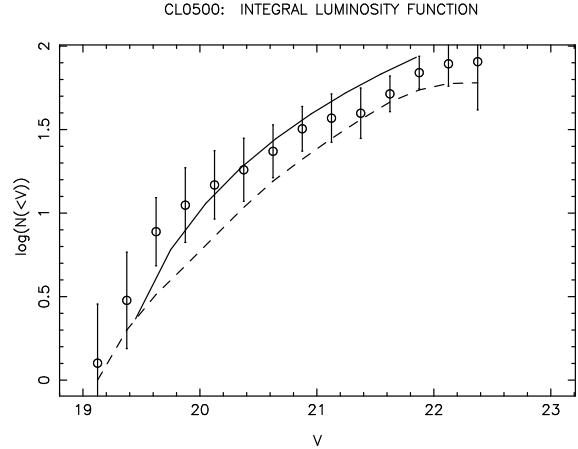

**Fig. 10.** CL0500-24 $V$ Luminosity Function. The dashed line represents the background counts and the solid line is a fit to the Schechter Function. We have adopted 22 as the completeness limiting magnitude. The error bars represent $\sqrt{N}$ noise

### 4.3. K – and evolutionary corrections

The values of $g^\star$ and $r^\star$ have to be corrected for evolution and K-correction. We use Bruzual's models to compute the sum of these two corrections. We have chosen to model an elliptical galaxy as the typical cluster galaxy at $M^\star$. The model parameters are those used by Bruzual & Charlot (1993) to reproduce the spectrum of NGC 3379, namely a unique burst of star formation at the formation redshift $z_f$, with a duration of 1 Gyr, and a Salpeter initial mass function. Other parameters are the Hubble constant, $q_o$ and the age of the Universe $(t_U)$. Adopted values are: $z_f = 4.34$, corresponding to a galaxy age of 10 Gyr, $H_o = 60$ km s$^{-1}$ Mpc$^{-1}$, $q_o = 0.5$ and $t_U = 10.88$ Gyr $(= 2/3H_o^{-1})$.

The corresponding sum of the K – and evolution corrections are: $M_g(z=0) - M_g(z=0.272) = -0.59$, $M_r(z=0) - M_r(z=0.272) = -0.11$ and $M_V(z=0) - M_V(z=0.322) = -0.50$, and the colours of such a galaxy at redshift 0 are, $(B-V) = 0.86$, $(g-r) = 0.52$. In fact, these colours are slightly incompatible with Kent's relations which, for $(g-r) = 0.52$, give $(B-V) = 1.07$. We adopt in the following the average value $(B-V) = 0.97$ for an elliptical galaxy at redshift 0.

Since our model is for an elliptical galaxy, internal absorption is neglected. The derived distance moduli for our adopted values of $H_o$ and $q_o$ are 40.79 (CL0017-20, $z = 0.272$) and 41.18 (CL0500-24, $z = 0.322$), respectively. The resulting absolute magnitudes of the SLF knee are listed in last column of Table ??. Transforming $g$ and $r$ values to $V$ band, using $(B-V) = 0.97$ and Kent's relations, and taking the unweighted mean of the results, we find $M_V^\star = -21.45$ for CL0017-20, compatible within the uncertainties with the value for CL0500-24 ($M_V^\star = -21.71$). Both values are well within the range of published values. A detailed comparison is beyond the present scope, and should take care of different hypothesis concerning exclusion of the brightest cluster member, $H_o$ and $q_o$ adopted values, and the value of $\alpha$.

The total luminosity of each cluster is derived integrating the Schechter luminosity function. The result is given by (Schechter 1976):

$$L_{\text{tot}} = \Gamma(0.75) n^\star L^\star = 1.2254 n^\star L^\star. \tag{12}$$

Uncertainties are derived from the corresponding uncertainties in $n^\star$ and $L^\star$. The precision of luminosities derived this way is about 30 %. Adopted solar absolute magnitudes are $M_B^\odot = 5.48$ and $M_V^\odot = 4.83$, which leads to $M_g^\odot = 4.91$ and $M_r^\odot = 4.80$, using Kent's relations. Total luminosities in the original bands are then:

CL0017-20: $L_g = 2.1 \pm 0.7 \times 10^{12}$ L$_\odot$
CL0017-20: $L_r = 2.7 \pm 0.8 \times 10^{12}$ L$_\odot$
CL0500-24: $L_V = 2.7 \pm 0.8 \times 10^{12}$ L$_\odot$

The values are hardly sensitive to variation of $\alpha$: a change of $\pm 0.15$ in $\alpha$ (from $-1.1$ to $-1.4$) changes $L$ by $\pm 8\%$. In order to compare luminosities and mass-to-light ratios, we also give the CL0017-20 luminosity in the $V$ band. To compute it, we convert $L_{\rm g}$ and $L_{\rm r}$ to $L_{\rm V}$, using the difference between solar and galaxy colours, and take the unweighted mean of both estimates. This gives:

CL0017-20: $L_{\rm V} = 2.3 \pm 0.5 \times 10^{12}$ $L_\odot$

Both clusters have similar luminosities.

### 4.4. Richness

We can estimate the richness of these clusters (in the sense of Abell's richness class) by counting the number of galaxies within an Abell radius and correcting for the background contribution. We can verify the results by integrating the Schechter luminosity function determined in the cluster core area, and extrapolating the result to the whole cluster area.

The most straightforward determination is obtained for CL0500-24, where our large CCD field covers more than an Abell radius. The Abell radius $r_{\rm A}$ (Abell 1958) is calculated as $1.717/z$ in arcmin, which corresponds to 524 pixels in the CL0500-24 field. Within this radius, 215 galaxies brighter than $V = 22$ are counted. Background contamination is calculated in an annulus between 700 and 800 pixels from the center, where it amounts to 82 galaxies. The number of cluster members within one Abell radius is then 65, corresponding to a richness class of 1.

Integrating the (background corrected) Schechter luminosity function between $m_3$ (19.33 in $V$) and $m_3 + 2$ gives 52 galaxies in the CCD area **A1**, whose radius is noted $r_{\rm CCD}$. As for the luminosity, this figure is hardly sensitive to changes in $\alpha$. Assuming a $r^{-2}$ density profile with a core radius $r_{\rm c}$ of 250 $h_{50}^{-1}$ kpc (Luppino et al. 1991), the correcting factor is given by:

$$N_{\rm A} = N \times \frac{\ln(1 + (r_{\rm A}/r_{\rm c})^2)}{\ln(1 + (r_{\rm CCD}/r_{\rm c})^2)}, \qquad (13)$$

which amounts to 1.28 and gives a number of cluster members within one Abell radius of 67, in good agreement with the previous figure.

For CL0017-20, the determination is complicated by the fact that the Abell radius (435 pixels) falls out of our CCD area. The maximum circular radius up to where we can count is about 150 pixels, within which we have 64 galaxies brighter than $g = 21.5$. The background contamination derived in §4.1 gives 19 galaxies in the same area, leaving 45 cluster members. Assuming again a $r^{-2}$ density profile with the same core radius gives a factor 1.95 for extrapolating to the Abell radius, and thus a richness of 88 galaxies, corresponding to Abell class 2.

The integration of the Schechter luminosity function between $m_3$ (19.41 in $g$) and $m_3 + 2$ gives 55 galaxies in the useful CCD area. The extrapolation to one Abell radius gives 87 galaxies, confirming the above figure.

These medium redshift clusters are thus less rich than CL 1358+6245 studied by Luppino et al. 1991. Note however that these authors overestimate its richness, because they extrapolate their counts to an Abell radius of 3 Mpc, which is the correct value at low redshift, but becomes 1.8 Mpc at the cluster redshift. Their corrected Abell richness should be 198 galaxies, in place of 231.

## 5. Masses and mass-to-light ratios

The method described in Gourgoulhon et al. (1992) has been applied to detect possible subconcentrations in the clusters and to determine the mass and mass-to-light ratio of both clusters. The resulting parameters are given in Table ??. Listed quantities are the crossing time, the collapse time, the virial radius, the one-dimensional velocity dispersion, the virial mass, the total luminosity in $V$ band, and the mass-to-visual light ratio. The Gourgoulhon et al. (1992) method was designed for low $z$ clusters. Since CL0017-20 and CL0500-24 are medium redshift clusters relativistic corrections are necessary. Table ?? gives the multiplicative corrections applied to the quantities in Table ??. Notice that the values of $\sigma_{\rm V}$ in Table ?? differ from the ones quoted in §3.2, both because of the relativistic corrections and because values in Table ?? are luminosity-weighted quantities.

Both the crossing time and the collapse time of the clusters are smaller than the age of the Universe, and their ratio (6.9 for CL0017-20 and 6.2 for CL0500-24) is close to the virial value ($2\pi$). Thus, the clusters are both bound and virialized. The virial mass is then a good estimator of the true mass. It is important to note that a preliminary value of $M/L = 820$ in CL0017-20 (Infante et al. 1992) is wrong, due to incorrect relative galaxy positions, resulting in an overestimated virial radius.

No subconcentration is detected in CL0017-20, as is apparent from the velocity histogram (Fig. ??). CL0500-24 appears to be composed of two subconcentrations. This is confirmed by the velocity histogram (Fig. ??). The northern one, called hereafter N subconcentration, is dominated by the pair of galaxies marked N and S. The southern one, called C subconcentration, is dominated by galaxy C. Members of each subconcentration are identified on Fig. ??; C subconcentration galaxies are labeled with a **c** and N subconcentration galaxies are labeled with a **n**. The virial mass of CL0500-24 listed in Table ?? is calculated taking both subcondensations as a whole. This is probably an overestimate of the mass. It is worth noting that splitting the cluster into two subconcentrations reduces the estimate of the total mass by a factor 2.2. Thus, if we calculate the mass of the cluster as the sum of the virial masses of each subcondensation then the mass-to-light ratio becomes 157 $M_\odot/L_\odot$.

Looking at Table ??, both clusters appear quite similar.

Table 5. Characteristics of both clusters

| Cluster | $t_{\rm cr}$ Gyr | $t_{\rm col}$ Gyr | $R_{\rm V}$ Mpc | $\sigma_{\rm V}$ kms$^{-1}$ | $M_{\rm V}$ $10^{14} M_\odot$ | $L_{\rm V}$ $10^{12} L_\odot$ | $M_{\rm V}/L_{\rm V}$ $M_\odot/L_\odot$ |
|---|---|---|---|---|---|---|---|
| CL 0017 | 0.15 | 1.0 | 0.26 | 888 | 2.9 | 2.3 | 127 |
| CL 0500 | 0.16 | 1.0 | 0.38 | 1309 | 9.1 | 2.7 | 337 |

\* All values have been corrected for relativistic effects.
\* Adopted cosmology: $q_{\rm o} = 0.5$, $H_{\rm o} = 60$ [km s$^{-1}$ Mpc$^{-1}$].
\* $R_{\rm V}$, $\sigma_{\rm V}$ and $M_{\rm V}$ are luminosity-weighted quantities (formulae are given in Gourgoulhon et al. 1992).

Table 6. Relativistic Corrections in Table ??.

| $t_{\rm cr}, t_{\rm col}$ | $R_{\rm V}$ | $\sigma_{\rm V}$ | $M_{\rm V}$ | $L$ |
|---|---|---|---|---|
| $\frac{D_L}{D}(1+z)^{-1}$ | $\frac{D_L}{D}(1+z)^{-2}$ | $(1+z)^{-1}$ | $\frac{D_L}{D}(1+z)^{-4}$ | $(\frac{D_L}{D})^2$ |

where $D$ is $\frac{cz}{H_o}$, and the luminosity distance is (Lang 1986):

$$D_L = D\left\{1 + \frac{z(1-q_{\rm o})}{(1+2q_{\rm o}z)^{1/2} + 1 + q_{\rm o}z}\right\} \qquad (14)$$

## 6. Discussion and comparisons

In this section, we wish to discuss possible systematic errors affecting our determination of mass-to-light ratios. We will also compare our results to the Coma cluster, usually used as a reference, and to other medium redshift clusters with $M/L$ determinations.

We first wish to compare our adopted luminosity-weighted virial mass with the unweighted virial mass and the projected mass (Bahcall & Tremaine 1981; Heisler et al. 1985). Our projected mass is calculated using the factor $32/\pi$, which corresponds either to test particles orbiting radially around a point mass, or to a self-gravitating system with an isotropic velocity distribution. Here again, the mass of CL0500-24 is estimated as the sum of its subconcentrations. For completeness, we also give the values corresponding to CL0500-24 as a whole. Results are given in Table ??

In the mean, unweighted virial masses are larger than weighted virial masses by 60 %, and projected masses (calculated with hypothesis described above) are larger than unweighted virial masses by 20 %.

One caveat of our determination of $M/L$ comes from the fact that different parts of the cluster are used to determine $M$ and $L$. Indeed, the luminosity is calculated integrating the Schechter luminosity function defined by all the galaxies within the CCD cluster area (after correction for background contamination). This corresponds to a radius of 1020 kpc in CL0500-24 and a region of 1050 x 1730 kpc in CL0017-20. In turn, velocity dispersions and virial radii, therefore masses, are calculated using only galaxies with measured velocities, which are contained within a smaller radius (about 600 kpc for both clusters). Assuming that our determination of velocity dispersion is correct all over the cluster, we have studied how the virial radius varies when the studied area increases. To do this, we have used all the galaxies with photometric information, randomly excluding a given number of background galaxies, according to the studied area (these galaxies must be uniformly distributed). We have first verified that galaxies with velocity provide a good sampling of the region where they lie. This is the case in both clusters. Then, we have computed the variation of $R$ in CL0500-24, when the radius increases from 600 to 1000 kpc. This gives an increase of $48^{+10}_{-22}$ %. The main uncertainty of this ratio comes from the uncertainty of the true number of background galaxies. The above error bars have been derived from a variation of a factor 2 in this number.

Table 7. Various mass estimators for both clusters

| Cluster | $M_V^w$ $10^{14} M_\odot$ | $M_V^{uw}$ $10^{14} M_\odot$ | $M_{proj}$ $10^{14} M_\odot$ |
|---|---|---|---|
| CL 0017 | 2.9 | 5.9 | 6.9 |
| CL 0500 C | 1.9 | 2.9 | 3.8 |
| CL 0500 N | 2.2 | 3.3 | 3.6 |
| CL 0500 C + N | 4.2 | 6.2 | 7.5 |
| CL 0500 whole | 9.1 | 12.4 | 14.6 |

In order to test how this influences our estimate of the mass-to-light ratio, we have built a data base of 165 galaxies within 1200 kpc from the centre of the Coma cluster, with velocities confirming their membership, and magnitudes from Godwin & Peach 1977 ($V_{25}$ system). The increase in virial radius is well confirmed amounting to 48.7 % between 600 and 1000 kpc. Moreover, as can be seen in table ??, the mass-to-light ration remains constant for radii larger than 500 kpc, up to 1200 kpc.

Table 8. Coma Cluster

| $^1R_{max}$ Mpc | $R_V$ Mpc | $L_{V_{25}}$ $10^{12} L_\odot$ | $\sigma_V$ kms$^{-1}$ | $M_V$ $10^{14} M_\odot$ | $^2 M_V/L_V$ $M_\odot/L_\odot$ | n |
|---|---|---|---|---|---|---|
| 0.267 | 0.217 | 1.3 | 813 | 2.0 | 155 | 42 |
| 0.503 | 0.327 | 2.0 | 883 | 3.5 | 176 | 78 |
| 0.695 | 0.466 | 3.0 | 1005 | 6.6 | 221 | 107 |
| 0.887 | 0.547 | 3.4 | 1026 | 8.0 | 233 | 121 |
| 1.066 | 0.603 | 3.7 | 983 | 8.1 | 217 | 131 |
| 1.300 | 0.716 | 4.4 | 945 | 8.0 | 202 | 142 |
| 1.385 | 0.738 | 4.5 | 943 | 9.2 | 202 | 147 |

* All values have been corrected for relativistic effects.
* Adopted cosmology: $q_\circ = 0.5$, $H_\circ = 60$ [km s$^{-1}$ Mpc$^{-1}$].
* $R_V$, $\sigma_V$ and $M_V$ are luminosity-weighted quantities.
[1] $R_{max}$ is the maximum radius within which the values are computed.
[2] $L_V = 1.5\ L_{V_{25}}$.

We therefore conclude that our mass-to-light ratio has probably to be increased by 50 %, at least in the case of CL0500-24. Nevertheless, due to the uncertain background correction, lack of knowledge of the variation of the velocity dispersion, we prefer keeping our raw values.

There are 253 clusters with measured redshifts larger than 0.18. Among them, only 14 clusters have more than 13 measured velocities. The derived velocity dispersions range from 632 to 2112 km s$^{-1}$. Only 3 clusters have measured $M/L$ ratios; these are: Abell 370 ($M/L = 165$ M$_\odot$/L$_\odot$ in blue and $M/L = 56$ M$_\odot$/L$_\odot$ in red, Mellier et al. 1988), Abell S910 ($M/L = 90$ M$_\odot$/L$_\odot$ in red, Sharples et al. 1985) and CL0017-20 ($M/L = 820$ M$_\odot$/L$_\odot$, Infante et al. 1992, but we have seen that this ultimate value is wrong).

In order to compare these values to ours, we have to reduce them to our adopted cosmology ($q_\circ = 0.5$, $H_\circ = 60$ km s$^{-1}$ Mpc$^{-1}$) and to the V band. This exercise gives for Abell S910: $M = 1.5 \times 10^{15}$ M$_\odot$, and $L_V = 2.2 \times 10^{13}$ L$_\odot$, thus $M/L = 68$ M$_\odot$/L$_\odot$. For Abell 370, we get $M = 1.1 \times 10^{15}$ M$_\odot$, and $L_V = 1.1 \times 10^{13}$ L$_\odot$, thus $M/L = 96$ M$_\odot$/L$_\odot$.

Masses are unweighted virial values. Compared to our clusters, masses are twice as large, and luminosities are 4 to 9 times as large.

The richness of Abell 370 is given by Abell et al. (1989) as $N_A = 40$ This is probably an incorrect value, because this is known as a very rich cluster. For ACO S910, Abell et al. give $N_A = 132$, and Sharples et al. give $N_A = 102$). In any case, these clusters are richer than ours, and their mass-to-light ratios appear smaller.

## 7. Summary and Conclusions

The present paper deals with the photometric and dynamical properties of two clusters of galaxies, namely CL0017-20 and CL0500-24. These two clusters seem alike in terms of dynamical properties, richness, compactness, luminosity function and both present arc-like structures.

The main results of the paper are as follows:

1. Magnitudes and colours were obtained for 213 galaxies in CL0017-20 and 759 galaxies in CL0500-24. Care was taken to correctly remove the background galaxies. Galaxy number counts in CL0500-24, region **A2**, are consistent with results in other studies. The colour-magnitude diagram of CL0500-24, when contrasted with the colour-magnitude diagram of the surrounding field, shows that most of blue objects in the central area are probably field galaxies and not B-O objects. Further spectroscopy is required for the blue objects in the central region.
2. The luminosity functions of the two clusters have been fitted by a Schechter function. The Schechter parameters $n^*$ and $M^*$ (74, -20.99 in $g$ for CL0017-20 and 52, -21.71 in $V$ for CL0500-24) are consistent with those obtained in other studies.
3. Velocities were obtained for 26 galaxies in CL0017-20 and 4 in CL0500-24. Moreover, we have used 24 extra velocities from Giraud 1990. The distribution of velocities in both clusters are very close to Gaussian, in particular if CL0500-24 is split into two subconcentrations. Observed mean redshifts are $<z>_{CL0017-20} = 0.272$ and $<z>_{CL0500-24} = 0.322$, and observed line-of-sight velocity dispersions are $\sigma_{CL0017-20} = 1510$ km s$^{-1}$ and $\sigma_{CL0500-24} = 1911$ km s$^{-1}$, or $\sigma_{CL0500-24C} = 917$ km s$^{-1}$ and $\sigma_{CL0500-24N} = 1152$ km s$^{-1}$, when CL0500-24 is split into its subconcentrations.
4. A group finding algorithm was applied to the data in both clusters. No subconcentration is found in CL0017-20. However, our algorithm clearly detects two subconcentrations in CL0500-24.
5. Evolutionary and K − corrections were applied to the luminosities of both clusters. After correcting all quantities for relativistic effects, we obtain $(M/L)_{CL0017-20} = 127$ M$_\odot$/L$_\odot$ and $(M/L)_{CL0500-24} = 157$ M$_\odot$/L$_\odot$. Although both clusters show conspicuous arc-like structures centered on the cluster, they have low mass-to-light ratios.
6. A determination of Abell richness class has been done. CL0017-20 appears to be class 2 (88 galaxies in one Abell radius) and CL0500-24 class 1 (65 galaxies).

*Acknowledgements.* We would like to thank G. Bruzual for providing the galaxy evolution code, and C.J. Pritchet for the C.T.I.O. 4m CCD image of CL0017-20. Part of this work was supported by Fondecyt projects 90/635, 93/570 and 92/754